\documentclass[a4paper,fleqnm]{cas-sc}

\usepackage[authoryear,longnamesfirst]{natbib}

\usepackage{amsmath}
\usepackage{hyperref,cases,subcaption}

\def\av#1{{\langle  #1 \rangle}}
\def\be{\begin{equation}}
\def\ee{\end{equation}}
\def\bea{\begin{eqnarray}}
\def\eea{\end{eqnarray}}
\def\bsn{\begin{subnumcases}}
\def\esn{\end{subnumcases}}

\def\bml{\begin{mathletters}}
\def\eml{\end{mathletters}}

\def\av#1{{\langle  #1 \rangle}}

\newcommand{\tc}{\textcolor}

\def\AD#1{\tc{black}{#1}}
\def\eq#1{(\ref{#1})}

\begin{document}
\let\WriteBookmarks\relax
\def\floatpagepagefraction{1}
\def\textpagefraction{.001}
\shorttitle{Polygenic trait at equilibrium}
\shortauthors{A. Devi et~al.}

\title [mode = title]{When can fitness epistasis be ignored in a polygenic trait at equilibrium?}

\author[1]{Archana Devi}[orcid=0000-0003-3597-7316]
\cormark[1]
\fnmark[1]
\ead{archana.devi@univie.ac.at}
\ead[url]{https://ufind.univie.ac.at/en/person.html?id=1010614}
\credit{Conceptualization, Formal analysis, Investigation, Methodology, Software, Visualization, Writing - review and editing}
\affiliation[1]{organization={Department of Mathematics, University of Vienna},
                addressline={Oskar-Morgenstern-Platz 1}, 
                postcode={1090}, 
                city={Vienna},
                country={Austria}}

\author[2]{Kavita Jain}[orcid=0000-0002-7406-4109]
\fnmark[2]
\ead{jain@jncasr.ac.in}
\ead[URL]{https://www.jncasr.ac.in/faculty/jain}
\credit{Conceptualization, Formal analysis, Investigation, Methodology, Supervision, Writing - original draft, Writing - review and editing}
\affiliation[2]{organization={Theoretical Sciences Unit, 
Jawaharlal Nehru Centre for Advanced Scientific Research,},
                addressline={Bangalore}, 
                postcode={560064}, 
                postcodesep={}, 
                city={},
                country={India}}

\begin{abstract}
\AD{Although many phenotypic traits are determined by a large number of genetic variants, the behavior of allele frequencies in a polygenic trait is not completely understood.} The problem is especially challenging when the quantitative trait of interest is under epistatic selection as the allele frequency at a locus is affected by those at other loci. 
Here, we consider a panmictic, diploid finite population evolving under stabilizing selection and symmetric mutations when the population is in linkage equilibrium. In the stationary state, using a diffusion theory, we calculate the marginal distribution of allele frequency, and find parameter regimes where fitness epistasis can not be ignored for an accurate description of the frequency distribution. 
For such parameters, the mean deviation in the phenotypic optimum and genic variance are, however, found to be well captured even when epistatic interactions are neglected.  
Thus, while the presence of epistasis may not be evident in phenotypic quantities, it can strongly affect the allele frequency distribution. 
We also find that the allele frequency distribution at a locus is unimodal if its effect size is below a threshold effect and bimodal otherwise; these results are the stochastic analog of the deterministic ones where the stable allele frequency becomes bistable when the effect size exceeds a threshold. 
Our analytical results are verified against Monte Carlo simulations and numerical integration of a Langevin equation. 
\end{abstract}




%



\maketitle

\section{Introduction}


The evolutionary dynamics of complex phenotypic traits have traditionally been studied   in the framework of quantitative genetics that deals with the summary statistics of a trait; 
however, a detailed understanding of the genetic basis of phenotypic variation is crucial for predicting the evolutionary outcomes of quantitative traits such as complex diseases \citep{GarciaGonzalez:2026}. Population genetics focuses on the genetic details of the phenotypic trait as 
the allele frequencies underlying a trait evolve under the joint action of evolutionary forces such as mutation, selection, random genetic drift and gene flow, and informs us on the evolution and maintenance of phenotypic variation in a population and between populations and species. Due to the advancement of genome-wide association studies (GWAS) in recent years, much information about the genetic architecture of phenotypic traits has increasingly become available, and  it is now known that  the phenotypic variation depends on the number of genetic variants underlying a phenotype, the effect size and frequency of these variants, the interaction between these genetic variants and with the environment, etc. \citep{Visscher:2017, Timpson:2018}. For these reasons, much work in the last two decades has focused on understanding quantitative traits at a genetic level and connecting these insights to the phenotype \citep{Rockman:2012,Jain:2017c,Sella:2019,Barghi:2020}.

The evolution of a quantitative trait in terms of the underlying allele frequencies is also theoretically challenging  \citep{Burger:2000,Walsh:2018} since stochastic models of polygenic adaptation are described by high-dimensional Fokker-Planck equations, and 
the phenotypic selection is typically epistatic, that is, it is nonlinear in the trait value and therefore can not be expressed as the sum of fitness of individual phenotypic values. Several recent studies have shown that phenotypic traits such as birth weight in humans are under stabilizing selection which has a tendency to reduce phenotypic variation in a population as the fitness of a trait value decreases with increasing deviation from the phenotypic optimum
\citep{Robertson:1956,Kingsolver:2001,Sanjak:2018,Villemereuil:2020,Sodeland:2022}. 
If the population is initially far from the optimum, one can describe the evolutionary dynamics assuming directional selection  which, however, lacks epistasis \citep{Goetsch:2024}. But at large times, when the population is close to the optimum or has reached mutation-selection-drift equilibrium, the epistatic nature of stabilizing selection - both at phenotypic level as also at genetic level due to phenotype-genotype map - comes into play, and it is important to understand if and how these epistatic interactions affect the allele frequency distribution and the phenotypic quantities.

\AD{Here, we consider a polygenic trait evolving under stabilizing selection and symmetric mutations in a panmictic finite population under linkage equilibrium, and focus on its equilibrium properties. Due to stabilizing selection, the allele frequencies interact with each other and evolve such that the trait value approaches the optimum. Although the joint distribution of allele frequencies in the stationary state is known \citep{Wright:1937, Kimura:1964}, the distribution is not transparent enough to give an insight into the genetic architecture of the phenotypic trait. In this study, we obtain the marginal distribution of allele frequencies and study its properties in detail.} Our main conclusion is 
that for a polygenic trait controlled by a large number of loci, as is relevant in many biologically realistic scenarios, the allele frequency distribution can be accurately described ignoring epistatic interactions, provided selection is sufficiently strong. But for weak to moderate selection, this is possible under certain conditions on mutation and selection parameters and if the locus under consideration has sufficiently small effect size. However, irrespective of selection strength, the genic variance is found to match well with Bulmer's expression \citep{Bulmer:1972} that has been obtained on neglecting epistasis. Our work also extends the previous theory for infinitely large populations \citep{Vladar:2014,Jain:2017a}  to large but finite populations. 



\section{Model}
\label{model}

We consider the classic Latter-Bulmer model \citep{Latter:1960,Bulmer:1972} that describes a panmictic, diploid population of size $N$ characterized by a single polygenic trait which is controlled by $L$ diallelic loci. 
The phenotype-genotype map is assumed to be additive \citep{Hill:2008}, and the trait value $z=\sum_{i=1}^L \frac{\gamma_i}{2} (\sigma_i+\sigma^*_i)$ where $\sigma_i, \sigma^*_i$, respectively, denote the maternal and paternal contribution to the trait of an individual and take the value $1 (-1)$ if {mutant (wildtype) allele} is present at the $i$th locus. Here the effect size of either allele is $\gamma_i$, and is chosen independently of the effect size at other loci from a common distribution $p(\gamma)$ which is assumed to be an exponential distribution with mean ${\bar \gamma}$ \citep{Mackay:2004,Goddard:2009}. 

The phenotypic trait evolves under stabilizing selection \citep{Robertson:1956,Kingsolver:2001,Sanjak:2018,Villemereuil:2020,Sodeland:2022} with the {phenotypic fitness function, } 
\be
w(z) \approx 1-\frac{s}{2} (z-z_o)^2 \label{stabsel}
\ee
which decays quadratically away from a (time-independent) optimal trait value $z_o$, and where $s$ denotes the strength of selection. 
We also assume that the mutation between wildtype and mutant allele occurs at an equal rate $\mu$ at each locus.

The dynamics follow a classical Wright-Fisher model where genotypes of the individuals undergo random genetic drift, selection, recombination, mutations, and random matings with other individuals to form the next generation - while  selection, recombination, and mutation are deterministic forces, the allele frequencies fluctuate in a finite population due to random genetic drift. At large times, the population always reaches a steady state under the joint action of these biological forces, and here we focus on the resulting equilibrium state. Under linkage equilibrium (see Appendix~\ref{app_le}), the genotype frequencies can be expressed as the product of allele frequency at each locus  in a fully recombining population; however, the allele frequencies do not evolve independently due to the indirect epistatic interaction imposed by the fitness function \eq{stabsel}. 

In the following sections, the model described above is studied analytically using a diffusion theory, and numerically using Monte Carlo (MC) simulations and by integrating a Langevin equation using Euler-Maruyama (EM) method. As discussed in Sec.~\ref{methods} and Sec.~\ref{var}, for a population in linkage equilibrium, the MC simulations are accurate for low mutation rates, but EM method works better for high mutation rates. For this reason, below we describe both methods.

\begin{figure}[t!]
 \includegraphics[angle=0, scale=0.5]{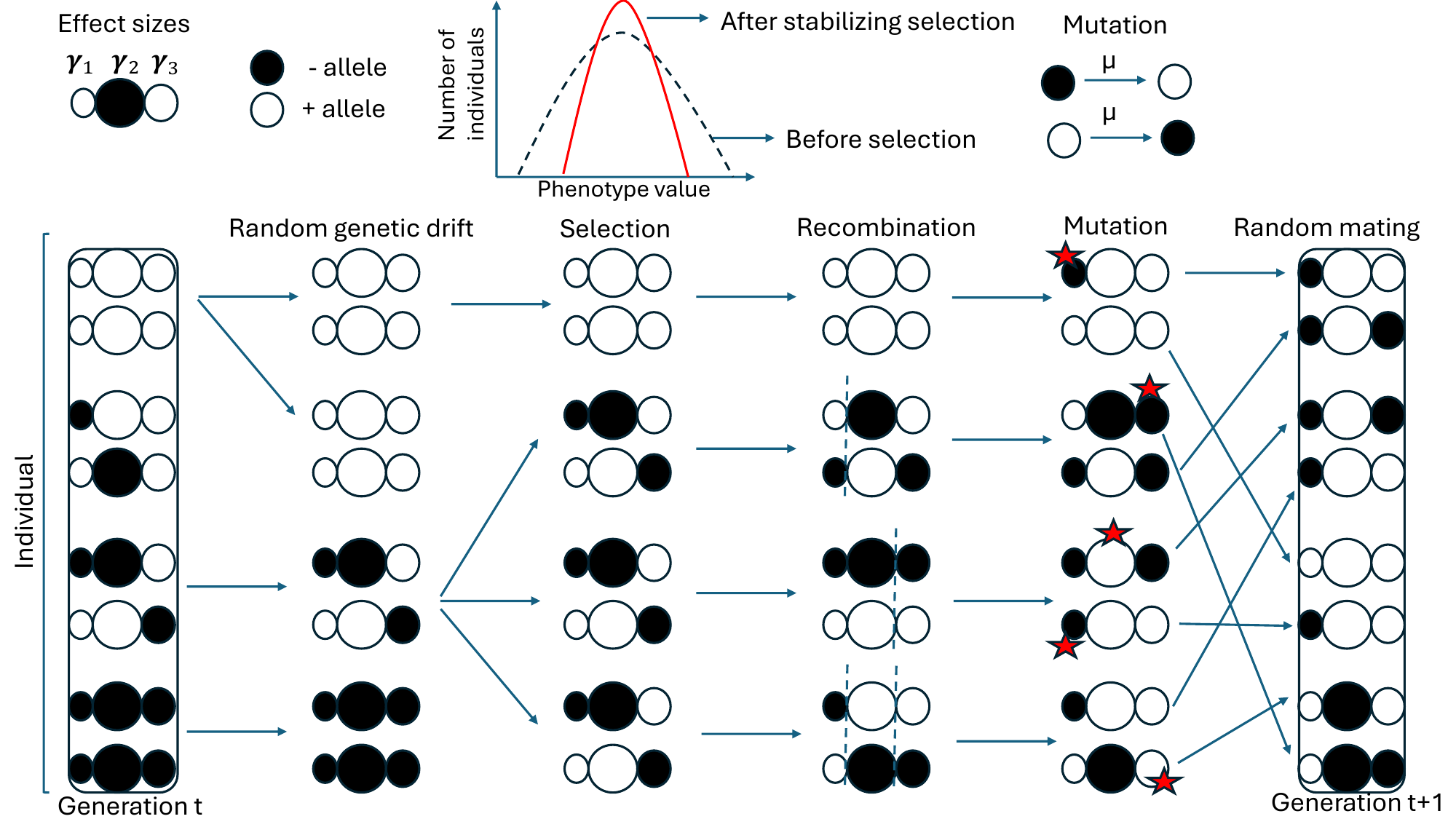}
 \caption{Diagram depicting how each evolutionary force shapes the genotypic configuration of individuals in a Wright-Fisher process implemented in  Monte Carlo simulations. Here, we consider $N=4$ diploid individuals  with $L=3$ loci to show the simulation steps from generation $t$ to $t+1$.}
 \label{fig_model}
\end{figure}

\section{Methods}
\label{methods}

The model dynamics  implemented in MC simulations are depicted in Fig. \ref{fig_model}. We consider a population of $N$ individuals, each having two chromosomes with $L$ sites and assign an effect value $\gamma$ to each locus (the effect sizes are exponentially distributed unless stated otherwise). Then each locus of a chromosome is multiplied by $+1$ or $-1$ with its effect size, where $+1$ or $-1$ are chosen with equal probability using a uniform random number generator. The trait value of each individual is half of the sum of its two chromosomal values. The population evolves in discrete and nonoverlapping generations 
 through the following steps that are performed at each generation:

\noindent(i) Random genetic drift: We choose individuals randomly from the population using a uniform random number generator. However, the viability of a particular individual depends on its relative phenotypic fitness keeping the total number of individuals constant every generation.  

\noindent(ii) Selection:  Each individual is assigned a fitness according to \eq{stabsel}. We weight the individual fitness with the maximum fitness in the population since the viability of the individual depends on its relative fitness in the population rather than its absolute fitness. We select the individual if a uniform random number is less than its relative fitness.

\noindent(iii) Recombination: Two chromosomes of the selected individual recombines according to the recombination parameter. Since we assume free recombination in this study, we find the recombination break points by drawing a Bernoulli-distributed random number with probability one half at every locus.  After finding all the breakpoints, we exchange the genetic material of the two chromosomes at each recombination breakpoint. After the recombination event, we keep either of the recombined chromosomes with equal probability and discard the other one for every individual. This process continues until we select $2N$ chromosomes. 

\noindent(iv) Mutation: We change the sign of the effect size of each locus if a standard uniform random number is less than the mutation probability. 

\noindent(v) Random mating: We get $N$ offspring for the next generation by pairing chromosomes randomly from the $2N$ chromosomes. 



As shown in Appendix~\ref{app_fpe}, under linkage equilibrium, the above model can be described using a diffusion theory. Following It{\^ o} prescription, corresponding to the Fokker-Planck equation \eq{LEfullfpe}, the Langevin equation for the allele frequency $x_i$ at the $i$th locus   is given by [refer to Sec.~4.3.5, \citet{Gardiner:1997}]
\be
dx_i(t)=\av{\delta x_i} dt+ \sqrt{\av{(\delta x_{i})^2}} \; dW_i(t) \label{LEdef}
\ee
where, as discussed in Appendix~\ref{app_fpe}, 
\bea
\av{\delta x_i} &=&  -\frac{s}{2} [2 \gamma_i x_i (1-x_i) \Delta c_1+  \gamma_i^2 x_i (1-x_i) (1-2 x_i)] 
+ \mu (1-2 x_i) \label{LEdetp} \\
\av{(\delta x_{i})^2} &=& \frac{x_i(1-x_i)}{2 N}
\eea
and $dW_i$ is the Wiener process. The Langevin equation \eq{LEdef} is integrated numerically using the EM method \citep{Higham:2001}.  Dividing the time $t$ into $t/\delta t$ intervals of equal length $\delta t$, the allele frequency at the $i$th locus is updated as (Sec.~4.3.1, \citet{Gardiner:1997})
\be
x_i(n+1)=x_i(n)+ \av{\delta x_i}  \delta t+  \sqrt{\av{(\delta x_{i})^2}}~\eta~,~0 \leq n \leq \frac{t}{\delta t}
\label{Langevin}
\ee
where $\eta$ is a random variable chosen independently at each time step from a normal distribution with mean zero and variance $\delta t$. In all the numerical data presented using the above equation, the population starts with allele frequency one half at every locus, and time step size $\delta t=0.1$. 

For either method, after a burn-in period of $10 N$ generations, the data in the stationary state are obtained 
by averaging over $10^5-10^8$ generations, depending on the parameters. 
Although the EM method is computationally faster and easier, as shown in Fig.~\ref{varMCEML},  it fails to accurately capture the genic variance (see Appendix~\ref{app_le} and Sec.~\ref{var}) for low mutation rates. The EM method is  designed to simulate continuously fluctuating, large-population, and can not correctly model the small population dynamics or low mutation rate where fixation or loss of alleles occur \citep{Hutzenthaler:2011}. On the other hand, the MC simulations which are computationally expensive for large populations with large number of loci, are accurate for low mutation rate; however, the linkage equilibrium  gets distorted when both mutation and selection are strong in a finite population which generates (and maintains) non-random associations between loci faster than the maximum recombination probability equal to one half can dissipate them. These non-random associations prevent the population from reaching the perfect linkage equilibrium in MC simulations, but the EM method uses the Langevin equation which is derived 
 assuming linkage equilibrium. For this reason, we use the appropriate method to obtain the numerical results discussed below.

\section{Steady state distribution}

In the steady state and under linkage equilibrium, the exact joint distribution of mutant allele frequencies has been obtained in the framework of diffusion theory, and  is given by  \citep{Kimura:1964} [also see Appendix~\ref{app_fpe}]
\bea
P(\vec x) \propto e^{-{N s (c_1-z_o)^2}}   \prod_{i=1}^L \psi_B(x_i) \label{LEPss1}
\eea
where, 
\be
c_1(\vec x)=\sum_{i=1}^L \gamma_i (2 x_i-1) \label{c1def}
\ee
is the population-averaged phenotype, and 
\be
\psi_B(x_i) \propto (x_i (1-x_i))^{4N\mu-1} e^{-2 N s \gamma_i^2 x_i (1-x_i)}  \label{indmarg}
\ee 
Note that due to  linkage equilibrium, although the population dynamics at each locus occur independently of the other loci, the distribution \eq{LEPss1} is not a product measure; this is because due to epistasis, the selection \eq{stabsel} is nonlocal and depends on all loci through the global variable $c_1(\vec x)$. 


\subsection{Marginal distribution for a large number of loci}

\begin{figure}[t!]
\begin{subfigure}{0.32\textwidth}
\caption{} 
\includegraphics[scale=0.44]{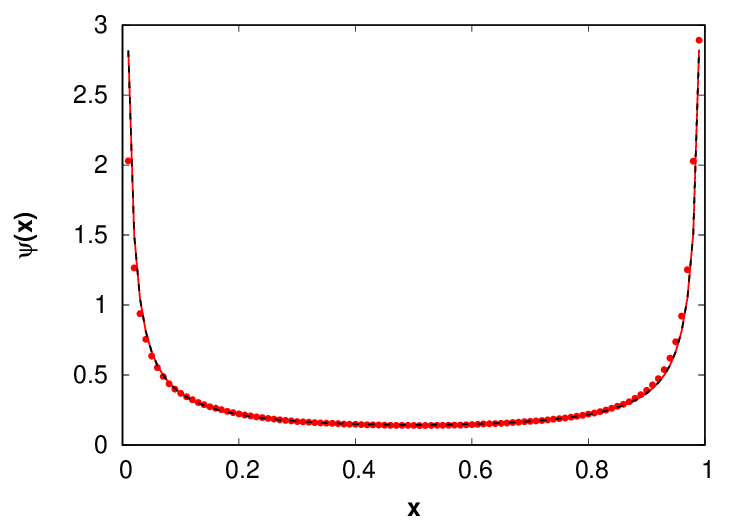}
\end{subfigure}
\begin{subfigure}{0.32\textwidth}
\caption{} 
\includegraphics[scale=0.44]{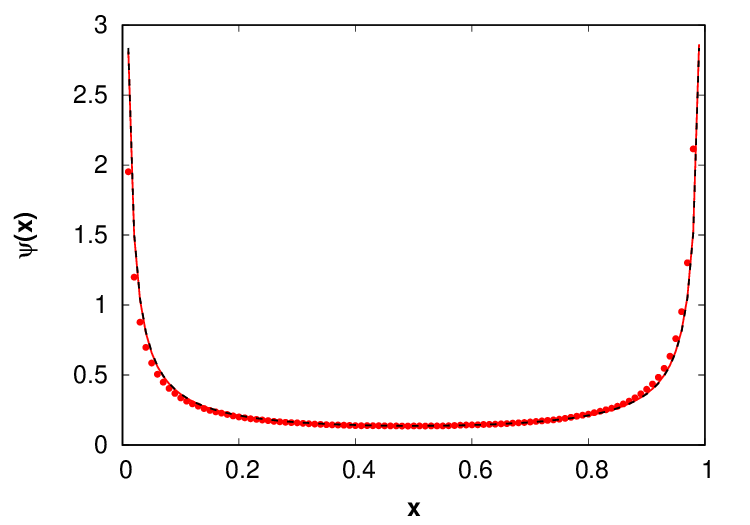}
\end{subfigure}
\begin{subfigure}{0.32\textwidth}
\centering
\caption{} 
\includegraphics[scale=0.44]{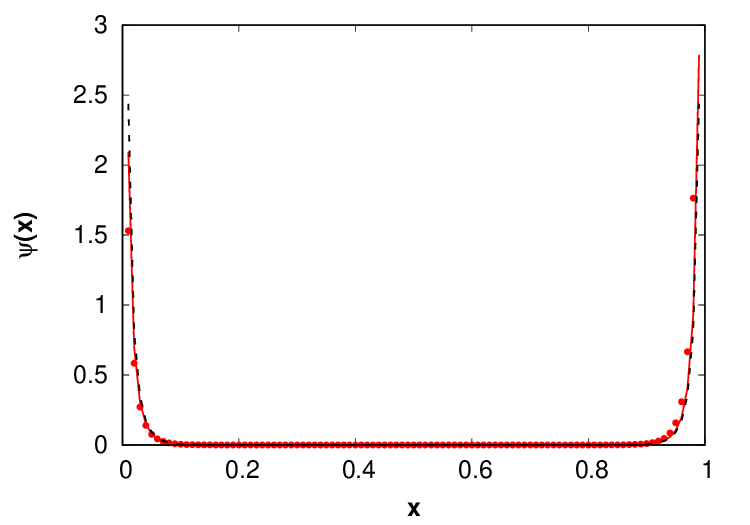}
\end{subfigure}
\caption{Marginal distribution for weak mutation ($4 N \mu < 1$) at a locus with effect size (a) $\gamma_i=0.01$, (b) $\gamma_i=0.05$ and (c) $\gamma_i=0.7$. The parameters are $N=1000$, $s=0.05$, $\mu=0.00002, L=200$, $\bar{\gamma}=0.1$, and  $z_o=1$. The deterministic threshold size ${\hat \gamma}=0.056$ for these parameters whereas  stochastically, there is no threshold effect. 
The points are obtained from MC simulations and the red solid line shows  the analytical expression (\ref{psimain}) where 
$\kappa_2 \approx 4.8$ for the effect sizes used in this plot. Here, $2 N s {\bar \gamma}^2=1$ and as both $\kappa_2$ and $2 N s \kappa_2$ are large, \eq{epiig} is satisfied and the marginal distribution (\ref{indmarg}) shown by black dashed line matches the distribution $\psi$.}
\label{FigSI_Uless1}
\end{figure}

To understand the allele frequency distribution at a locus, we now consider the marginal distribution $\psi(x_i)$ of the frequency at the $i$th locus which can be obtained by integrating the joint distribution $P(\vec x)$ over all the allele frequencies except $x_i$ and using the central limit theorem. For a large number of loci, as shown in Appendix \ref{app_margdistr}, the single-locus distribution can be approximated  by 
\be
\psi(x_i) \propto (x_i (1-x_i))^{4N\mu-1} e^{-2 N s \gamma_i^2 x_i (1-x_i)} e^{\frac{-N s [ \gamma_i (2 x_i-1)-z_o]^2}{1+2 N s \kappa_{2,i}}} 
\label{psimain}
\ee 
{where,} 
\be
\kappa_{2,i} =  \sum_{j \neq i} 4 \gamma_j^2 (\av{x_j^2}_{\psi_B}-\av{x_j}_{\psi_B}^2) \label{k2pdefn}
\ee 
is the {\it statistical variance} of allele frequency $x_i$ w.r.t. the distribution $\psi_B$ and given by \eq{app_k2i}. {As each summand on the right-hand side (RHS) of (\ref{k2pdefn}) is positive, $\kappa_{2,i}$ increases linearly with $L$ and captures the effect of epistatic interactions, as explained below.} 

 For $2 N s \kappa_{2,i} \gg 1$, the last factor in \eq{psimain} reduces to $e^{-\frac{[ \gamma_i (2 x_i-1)-z_o]^2}{2 \kappa_{2,i}}}$, and tends to unity provided $\kappa_{2,i}$ is also large (and $z_o$ does not scale linearly with $L$). Thus, 
 when $\kappa_{2,i} \gg 1$ \emph{and} $2 N s \kappa_{2,i} \gg 1$, the effect of epistatic interactions can be ignored and the marginal distribution \eq{psimain} reduces to $\psi_B(x_i)$:
 \be
 \psi(x_i) \stackrel{\kappa_{2,i} \gg 1, 2 N s \kappa_{2,i} \gg 1}{\longrightarrow} \psi_B(x_i) \label{epiig}
 \ee
On approximating $\kappa_{2,i}$ by $\kappa_2= \sum_{j=1}^L 4 \gamma_j^2 (\av{x_j^2}_{\psi_B}-\av{x_j}_{\psi_B}^2)$, as shown in Appendix~\ref{app_sve}, we obtain
\bsn
{\kappa_2 \approx \label{appm_gvar}}
\frac{2 L{\bar \gamma}^2}{1+8 N \mu}  &,~$N s {\bar \gamma}^2 \ll 2 $ \label{appm_gvar1} \\
2 L {\bar \gamma}^2  &,~ $N s  {\bar \gamma}^2 \gg 2$ \label{appm_gvar2}
\esn
For ${\bar \gamma} \sim 1$, when selection is strong ($2 N s {\bar \gamma}^2 \gg 1$), it is sufficient to ensure that $\frac{\kappa_2}{{\bar \gamma}^2}$ is also large in order to approximate the marginal distribution by $\psi_B$. Then \eq{appm_gvar2} shows that the epistatic factor in the marginal distribution can be ignored for large number of loci. However, for weak selection, $\kappa_2$ given by \eq{appm_gvar1} must be much larger than $(2 N s)^{-1}$ in order that epistasis can be neglected. 


For large $L$, the exponential on the RHS of \eq{psimain} can be expanded in powers of $L^{-1}$; however, as explained in \cite{Blinnikov:1998}, corrections to central limit theorem are required to obtain the correct expression to leading order in $L^{-1}$, and is briefly described in Appendix~\ref{app_edge}. 
 

\subsection{Transition in marginal distribution for strong mutation}
\label{transi}

In Fig.~\ref{FigSI_Uless1}, we show the marginal distributions obtained in simulations and in Fig.~\ref{FigSI_mostLE} and Fig.~\ref{margssnz} by solving the Langevin equation (\ref{Langevin}) numerically, and compare them with the analytical expression (\ref{psimain}). For weak mutation ($4 N \mu < 1$), as Fig.~\ref{FigSI_Uless1} shows, the marginal distribution (\ref{psimain}) is $\textrm{U}$-shaped, as expected. For the parameters in this figure, \eq{appm_gvar1} yields $\kappa_2 \approx 3.4, 2 N s \kappa_2 \approx 344$ so that \eq{epiig} is satisfied, and we observe a good match with $\psi_B(x_i)$. 

For strong mutation ($4 N \mu >1$), the distribution $\psi(x_i)$ has the following interesting property: it is unimodal if the effect size is below a threshold effect size ${\hat \gamma}_N(L)$ (see Sec.~\ref{thr}) and bimodal otherwise.  This is displayed in Fig.~\ref{FigSI_mostLE} where the frequency distribution at a locus with sufficiently small effect size has one peak and it is flat for intermediate frequencies at a locus with the threshold size (inflection point), while at a locus with large effect size, the distribution has two maxima which are quite symmetric about frequency one half for parameters in Fig.~\ref{FigSI_mostLE} and asymmetric in Fig.~\ref{margssnz}. These behavior are also reminiscent of second order phase transitions in physical or chemical systems where the potential energy (given here by $-\ln \psi$) changes as an external parameter such as temperature (here, $2 N s \gamma_i^2$) is varied. 
For the parameters in Fig.~\ref{FigSI_mostLE}, as $N s {\bar \gamma}^2=49$, using \eq{appm_gvar2}, we find that $\kappa_2 = 196$ which is comparable to value obtained using the effects used in this figure, and as selection is strong and $\kappa_2$ is large, we find that $\psi \approx \psi_B$; on the other hand, for the parameters in Fig.~\ref{margssnz}, since $N s {\bar \gamma}^2=0.64$, \eq{appm_gvar1} yields $\kappa_2 = 0.75$ and as discussed above, the marginal distribution $\psi$ is poorly estimated by $\psi_B$.

In Fig.~\ref{FigSI_mostLE} and Fig.~\ref{margssnz}, 
we also note a mismatch between the numerical data and the analytical expression \eq{psimain} when the locus has an effect size close to or above the threshold size. This is because the population spends a long time near one of the maxima before traversing the valley separating the other maximum \citep{Barton:1987b}. {Due to such shifts in the equilibria for large-effect loci, the numerical results for the stationary state distribution were obtained by averaging over $10^3$ independent initial conditions (ensemble-averaging) as well as long time periods in the stationary state (time-averaging), 
as this allowed us to sample the distribution near both the allele frequency peaks efficiently. 

\begin{figure}[t!]
\begin{subfigure}{0.23\textwidth}
\caption{} 
\includegraphics[scale=0.33]{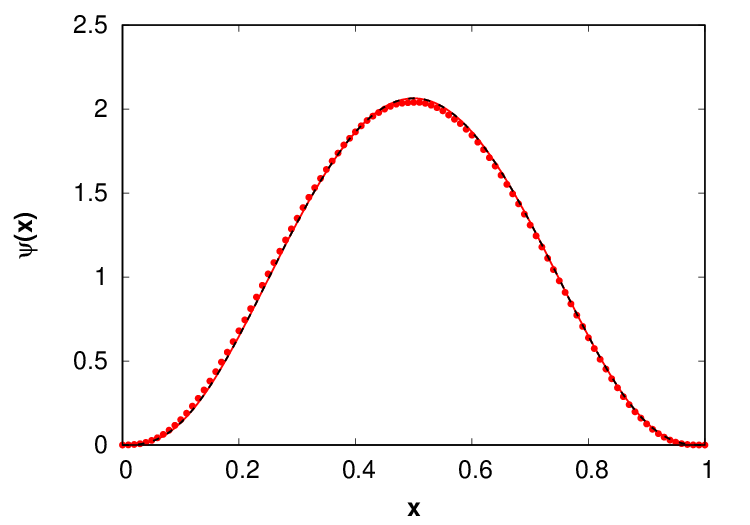}
\end{subfigure}
\begin{subfigure}{0.23\textwidth}
\caption{} 
\includegraphics[scale=0.33]{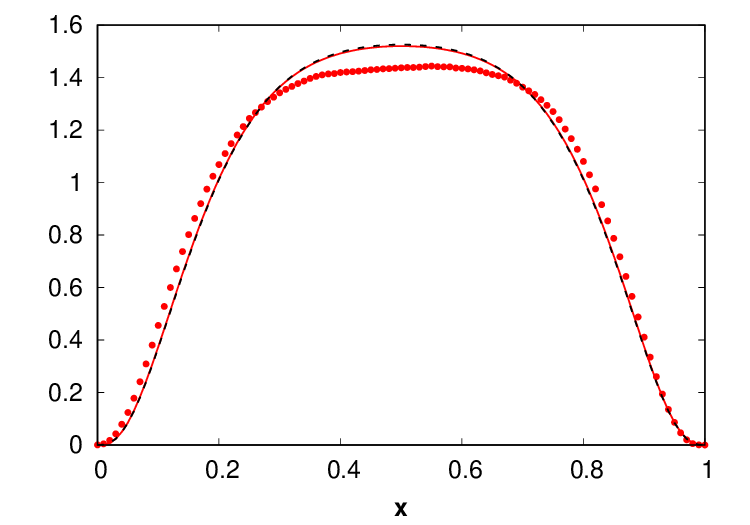}
\end{subfigure}
\begin{subfigure}{0.23\textwidth}
\caption{} 
\includegraphics[scale=0.33]{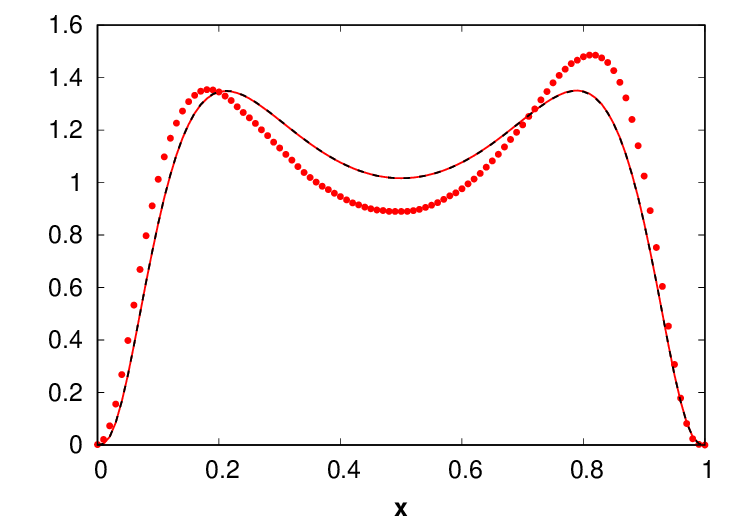}
\end{subfigure}
\begin{subfigure}{0.23\textwidth}
\centering
\caption{} 
\includegraphics[scale=0.33]{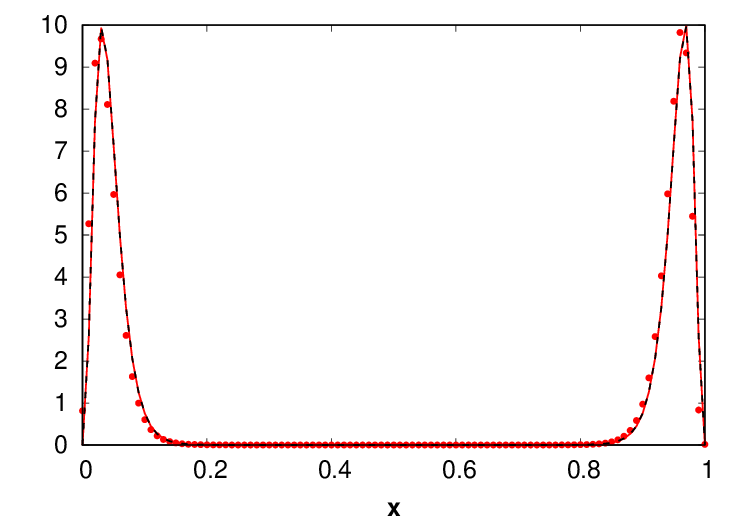}
\end{subfigure}
\caption{Marginal distribution for strong mutation ($4 N \mu > 1$) and strong selection ($N s {\bar \gamma}^2  > 2$) at a locus with effect size (a) smaller ($\gamma_i=0.1$), (b) {just below ($\gamma_i=0.23$)}, (c) just above ($\gamma_i=0.3$) and (d) larger ($\gamma_i=0.7$) than the threshold size ${\hat \gamma}_N(L)=0.24$. The parameters are $N=1000$, $s=0.1$, $\mu=0.001, L=200$, $\bar{\gamma}=0.7$, and  $z_o=1$. For the set of effects used here, there were $137$ large-effect loci. The points are obtained by solving (\ref{Langevin}) numerically, and the red solid line shows the analytical expression (\ref{psimain}) where $\kappa_2 \approx 235$ for the  effect sizes used in this plot. Here, as both 
$\kappa_2$ and $2 N s \kappa_2$ are large, \eq{epiig} is satisfied and the marginal distribution (\ref{indmarg}) shown by black dashed line matches the distribution $\psi$.}
\label{FigSI_mostLE}
\end{figure}

\begin{figure}[t!]
\begin{subfigure}{0.32\textwidth}
\caption{} 
\includegraphics[scale=0.44]{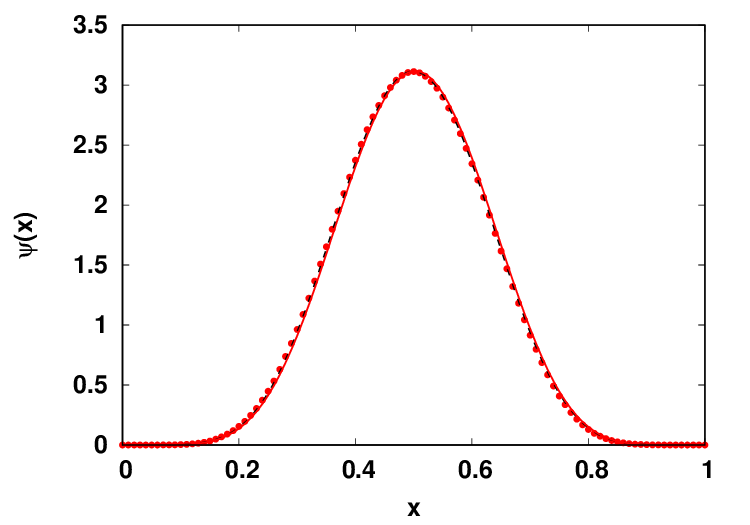}
\end{subfigure}
\begin{subfigure}{0.32\textwidth}
\caption{} 
\includegraphics[scale=0.44]{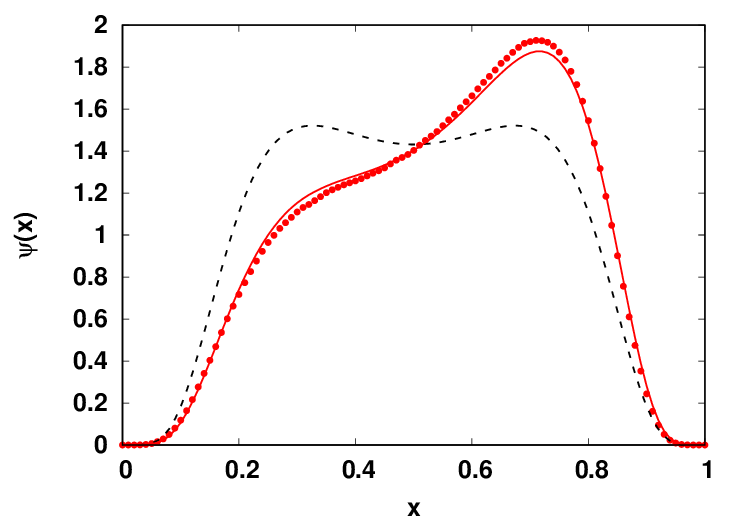}
\end{subfigure}
\begin{subfigure}{0.32\textwidth}
\centering
\caption{} 
\includegraphics[scale=0.44]{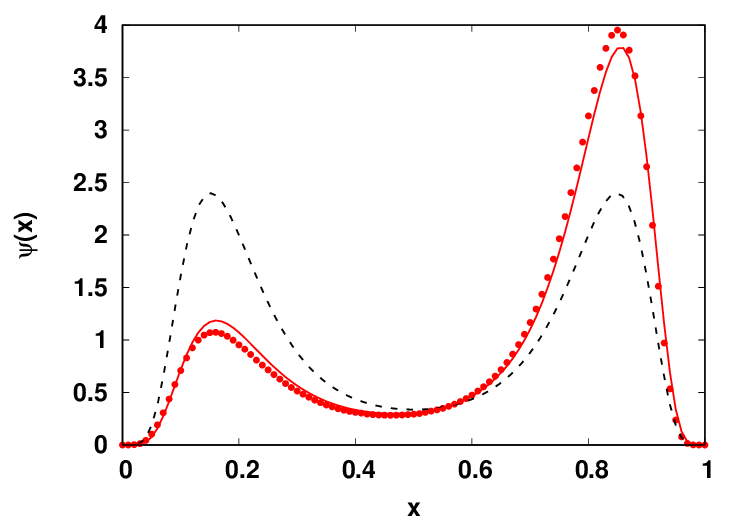}
\end{subfigure}
\caption{Marginal distribution for strong mutation ($4 N \mu > 1$) and weak selection ($N s {\bar \gamma}^2  < 2$) at a locus with effect size (a) smaller ($\gamma_i=0.05$), (b) close to ($\gamma_i=0.4$) and (c) larger ($\gamma_i=0.5$) than the threshold size ${\hat \gamma}_N(L) \approx 0.42$. The parameters are $N=1000$, $s=0.1$, $\mu=0.002, L=1000$, $\bar{\gamma}=0.08$, and  $z_o=2$. For the set of effects used here, there were $7$ large-effect loci. The points are obtained by solving (\ref{Langevin}) numerically, and the red solid line shows the analytical expression (\ref{psimain}) where $\kappa_2 \approx 1.24$  for the  effect sizes used in this plot. Here, although $2 N s \kappa_2 \approx 248$ is large, as $\kappa_2$ is not very large, \eq{epiig} is not satisfied and the marginal distribution (\ref{indmarg}) shown by black dashed line does not match the distribution $\psi$ at loci except when the effect size is small.}
  \label{margssnz}
  \end{figure}

\subsection{Phenotypic mean deviation} 
\label{app_mean}

Using the joint distribution \eq{LEPss1}, we find that the distribution of phenotypic mean \eq{c1def}  is given by
\bea
Pr(c_1) 
&\propto& e^{-N s (\Delta {c_1})^{2}}  \int_0^1...\int_0^1   \delta (c_1-\sum_{i=1}^L  \gamma_i (2 x_i-1)) \prod_{i=1}^L dx_i \; {\psi_B}(x_i)   \\
&\propto& e^{-N s  (\Delta {c_1})^{2}} e^{-\frac{c_1^{2}}{2 \kappa_2}} 
\eea
where $\Delta c_1=c_1-z_o$, $\kappa_2$ is given by (\ref{app_svar}) and the last expression is obtained using the central limit theorem for large $L$ \citep{Bulmer:1972,Lande:1976}. 
The above distribution gives the average and variance of the phenotypic mean to be 
\bea
\langle c_1 \rangle&=&\frac{2Ns \kappa_2 z_o}{1+2Ns\kappa_2} \approx z_o \left(1-\frac{1}{2 N s \kappa_2}+{\cal O}(L^{-2}) \right) \label{zmfull}\\
\langle (c_1-\langle c_1 \rangle)^2 \rangle &=& \frac{\kappa_2}{1+2Ns\kappa_2}  \approx  \frac{1}{2 N s} -\frac{1}{4 N^2 s^2 \kappa_2}+{\cal O}(L^{-2})\label{zvfull}
\eea
Then the deviation in the trait mean can be written as
\be
\langle {\Delta c_1}\rangle\approx - \frac{z_o}{2Ns\kappa_2}
\label{meandev}
\ee
From (\ref{zmfull}) and (\ref{appm_gvar}),  we find that if the magnitude of the phenotypic optimum does not increase with the number of loci, the deviation in the mean phenotype vanishes with increasing $L$ and the population is perfectly adapted (on an average). Intuitively, as the variance of mean becomes smaller with increasing $L$, the distribution of phenotypic mean gets 
narrower so that the first factor in \eq{psimain} (although still random) can be neglected. 

\subsection{Mean genic variance}
\label{var}

As discussed in Appendix~\ref{app_le}, the genic variance within the population is defined as 
\be
c_2=2 \sum_{i=1}^L \gamma_i^2 x_i (1-x_i) \label{gvdef}
\ee
where the mutant allele frequencies are random variables distributed according to \eq{LEPss1}. For equal effects and $z_o=0$, \citet{Bulmer:1972} has obtained an expression for the expectation value of the genic variance by arguing as follows: 
from \eq{LEPss1}, the expected genic variance can be rewritten as
\bea
\av{c_2} &=& 2 \gamma^2 
\int_0^1 dx_1... dx_L \;   Y e^{-4 N s \gamma^2 X^2} {\hat P}(\vec x) \label{Ec2def}
\eea
where $X=\sum_{i=1}^L  x_i-\frac{1}{2}, Y=\sum_{i=1}^L x_i(1-x_i)$ and ${\hat P} \propto \prod_{i=1}^L [x_i (1-x_i)]^{4 N\mu-1} e^{-2 N s \gamma_i^2 x_i (1-x_i)} $. Then, since $X$ is antisymmetric about $x_i=1/2$ while $Y$ and ${\hat P}$ are symmetric, it follows that the expectation value of $XY$ with respect to ${\hat P}$ is zero and it is concluded that ``Hence $X$ and $Y$ must also be uncorrelated'' \citep{Bulmer:1972}. This reasoning then gives  
\bea
\av{c_2}=\av{c_2}_{\psi_B}=2 \gamma^2 \int_0^1 dx_1... dx_L \;   Y  {\hat P}(\vec x)=2 L \gamma^2 \frac{2 \mu  N \, _1F_1\left(\frac{1}{2};4 \mu  N+\frac{3}{2};\frac{1}{2} \gamma ^2 N
   s\right)}{(8 \mu  N+1) \, _1F_1\left(\frac{1}{2};4 \mu  N+\frac{1}{2};\frac{1}{2} \gamma
   ^2 N s\right)} \label{varB}
\eea
where ${_1}F_1(a, b, z)$ is the confluent hypergeometric function [Chapter 13, \citep{DLMF}]. However, the exponential factor in the integrand on the RHS of \eq{Ec2def} is an even function of $X$, and on expanding the exponential in a power series in $X$, we immediately find that the expectation value of $Y e^{-4 N s \gamma^2 X^2} $ and $Y$ w.r.t. ${\hat P}$ are not equal, contrary to Bulmer's argument above. 

\begin{figure}[t!]
\begin{subfigure}{0.48\textwidth}
\caption{} 
\includegraphics[scale=0.65]{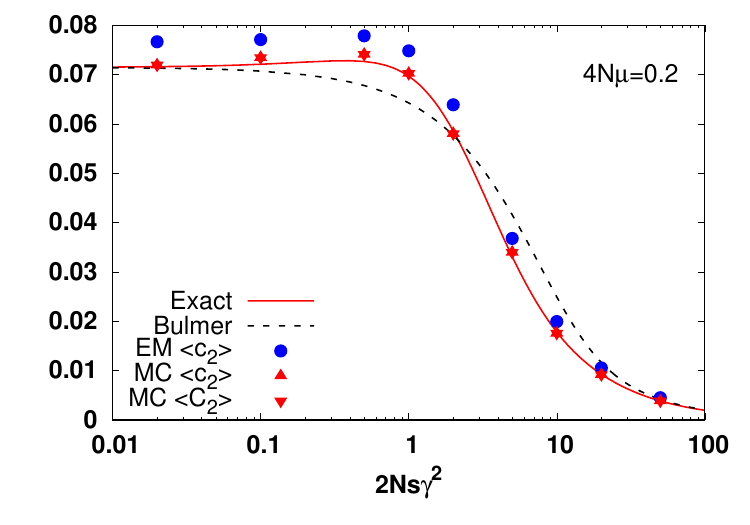}
\end{subfigure}
\begin{subfigure}{0.48\textwidth}
\caption{} 
\includegraphics[scale=0.65]{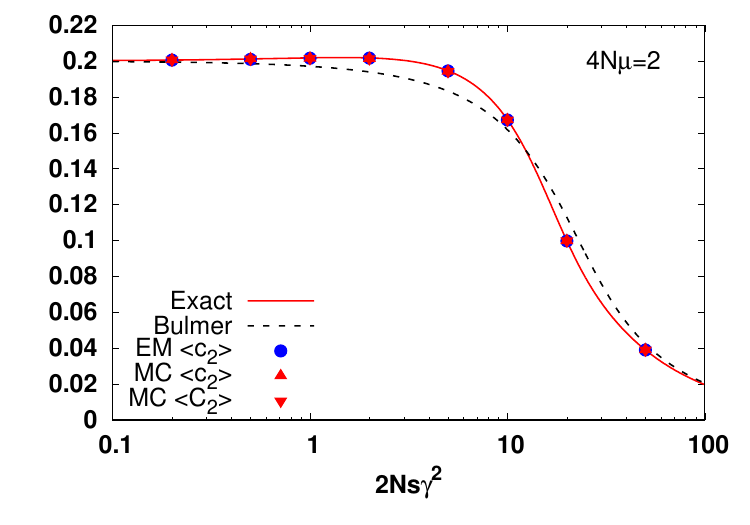}
\end{subfigure}
\caption{Mean genic variance \eq{gvdef} for $L=2$ for (a) weak and (b) strong mutation. In all the figures, $N=1000, z_o=0, \gamma_i=0.5$ (equal effects). In the legend, the line denoted by 'Exact' is obtained by numerically integrating the double integral of the joint
distribution in the result \eq{Ec2def} of diffusion theory, whereas 'Bulmer' shows the result \eq{varB} obtained by Bulmer, while 'MC' and 'EM', respectively, represent the data obtained using Monte-Carlo simulations and numerical integration of Langevin equation via Euler-Maruyama method.}
\label{varMCEM}
\end{figure}

For two loci, the integrals in \eq{Ec2def} can be readily done numerically and the expected variance is shown in Fig.~\ref{varMCEM}a and Fig.~\ref{varMCEM}b, and compared with $\av{c_2}_{\psi_B}$. 
We find that Bulmer's expression \eq{varB} does not match the exact result 
except for very weak selection (neutral limit) where the fitness epistasis is negligible and for very strong selection where variance in mean deviation given by \eq{zvfull} is negligible so that the loci evolve independently.  
These results are also compared with those obtained numerically. We find that the mean genic variance $\av{c_2}$ and mean genetic variance $\av{C_2}$ (where $C_2$ is defined in \eq{app_genvdef}) obtained in MC simulations are equal since the linkage disequilibrium is negligible for weak to moderate mutation rate with small number of loci where the recombination probability of $0.5$ can maintain the linkage equilibrium. The results from MC simulations are also in agreement with the exact results for both weak and strong mutation; however, the corresponding results for genic variance $\av{c_2}$ from EM method do not match the exact results for weak mutation for the reasons discussed in Sec.~\ref{methods}. 

In Fig.~\ref{varMCEML} where genic variance and genetic variance are shown for large $L$, we find that the mean genic variance is close to $\av{c_2}_{\psi_B}$. However, the mean genic variance and the mean genetic variance in MC
simulations are not equal for strong selection and strong mutation since the interference between genotypes is strong in this regime leading to a negative linkage disequilibrium between loci, and recombination with probability one half can not break all the associations between the loci; 
this results in a reduction in the genetic variance (\cite{Bulmer:1971,
Bulmer:1974}) which 
depends on the genic variance $\av{c_2}$ and the fitness variance $1/s$ (\cite{Turelli:1988,Turelli:1990,Turelli:1994}). 
Figure \ref{varMCEML} also shows that the EM method does not work for weak mutations, but it is in closer agreement with  Bulmer's expression than the MC simulation data when mutation is strong. 
From \eq{psimain},  we find that for large $L$, the corrections to Bulmer's result are of order $L^{-1}$, and these are discussed in Appendix~\ref{app_var}. 


\begin{figure}[t!]
\begin{subfigure}{0.48\textwidth}
\caption{} 
\includegraphics[scale=0.7]{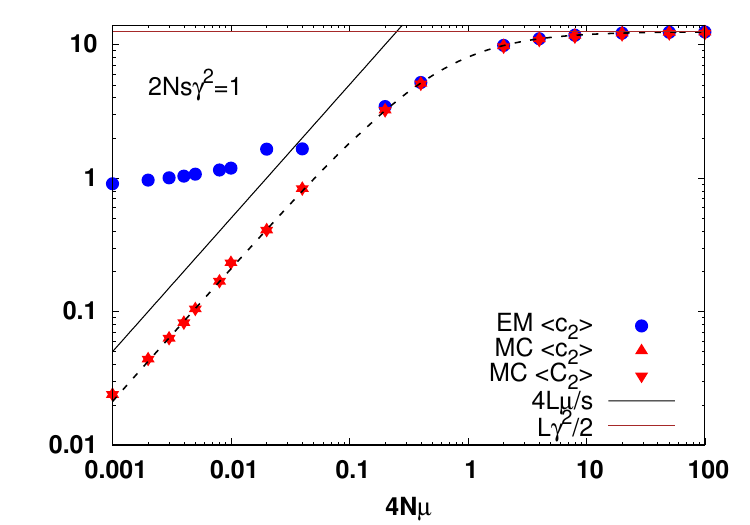}
\end{subfigure}
\begin{subfigure}{0.48\textwidth}
\caption{} 
\includegraphics[scale=0.7]{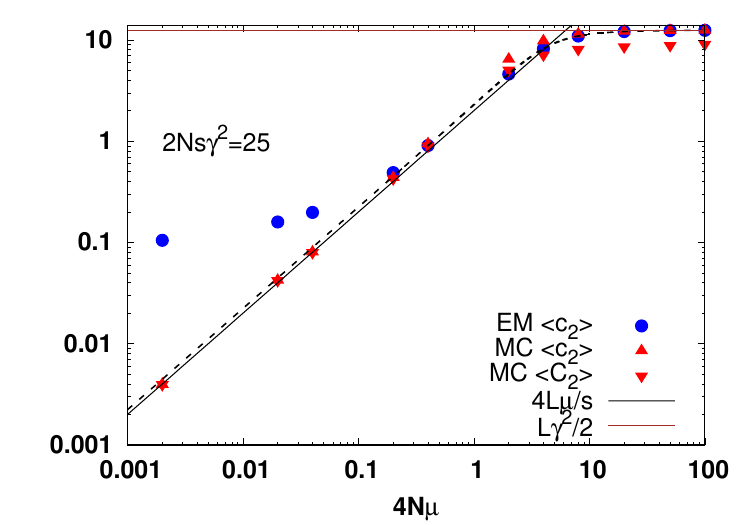}
\end{subfigure}
\caption{Mean genic variance \eq{gvdef} and mean genetic variance \eq{app_genvdef} for $L=100$ for (a) weak and (b) strong selection. In all the figures, $N=1000, z_o=0, \gamma_i=0.5$ (equal effects). In the legend, the 'MC' and 'EM', respectively, represent the data obtained using Monte-Carlo simulations and numerical integration of Langevin equation via Euler-Maruyama method. The black dashed line shows the result \eq{varB} obtained by Bulmer, while the black solid line shows the genic variance in the rare allele approximation \citep{Barton:1986}, and the brown line shows the genic variance {corresponding to all frequencies equal to one half} in the strong mutation limit. }
\label{varMCEML}
\end{figure}


\subsection{Threshold effect size}
\label{thr}

We finally describe the threshold effect size introduced in Sec.~\ref{transi}, and how it depends on $L$ and $N$. 
For $L \to \infty$, taking the derivative of the distribution \eq{indmarg} w.r.t. $x_i$, we immediately find that for $4 N \mu > 1$, the allele frequency distribution has one maximum if the effect size at a locus is below the threshold effect size ${\hat \gamma}_N$ and two maxima otherwise, where ${\hat \gamma}_N$ is given by 
\be
{\hat \gamma}_N=\sqrt{\frac{8\mu}{s}-\frac{2}{Ns}}=\sqrt{\frac{2}{N s} (4 N \mu-1)}
\label{gN}
\ee
The maxima of the allele frequency distribution (\ref{indmarg}) occur at
\begin{subnumcases}
{{\chi_i}=\label{tmm}}
\frac{1}{2} ~&,~$\gamma_i < {\hat \gamma}_N$\\
\frac{1}{2} \left(1\pm \sqrt{1-\frac{{\hat \gamma}_N^2}{\gamma_i^2}} \right) ~&,~$\gamma_i > {\hat \gamma}_N$
\label{tm}
\end{subnumcases}
and for $\gamma_i > {\hat \gamma}_N$, the maxima of the bimodal distribution are separated by a minimum at frequency $1/2$. 
{Thus, (\ref{gN}) and (\ref{tmm}) show that when selection is 
weaker than mutation ($\frac{N s \gamma_i^2}{2} \ll 4 N \mu-1$) or equivalently, for {\it small effect size} ($\gamma_i \ll {\hat \gamma}_N$), due to a high degree of polymorphism, the allele frequency distribution peaks at one half; however, when selection is stronger than mutation or for {\it large effect size}, the allele frequency distribution peaks at frequencies away from one half but the population is not monomorphic (unlike the $\textrm{U}$-shaped distribution in weak mutation  regime).}

The finite $L$ corrections to $\hat{\gamma}_N$ and $\chi_i$ are derived in Appendix \ref{app_modes}, 
and shown in Fig.~\ref{peakell}; we find that the modes of the allele frequency distribution at a locus with effect size away from the threshold effect ${\hat \gamma}_N(L)$ are well approximated by the result (\ref{tmm}) for infinite loci. But close to the threshold effect, they are substantially different; in particular, for positive (negative) phenotypic optimum, the maximum in the frequency distribution of the small-effect locus increases (decreases) with the effect size and occurs at a frequency which is substantially larger (smaller) than one half.

\begin{figure}[t!]
\begin{subfigure}{0.48\textwidth}
\caption{} 
\includegraphics[scale=0.5]{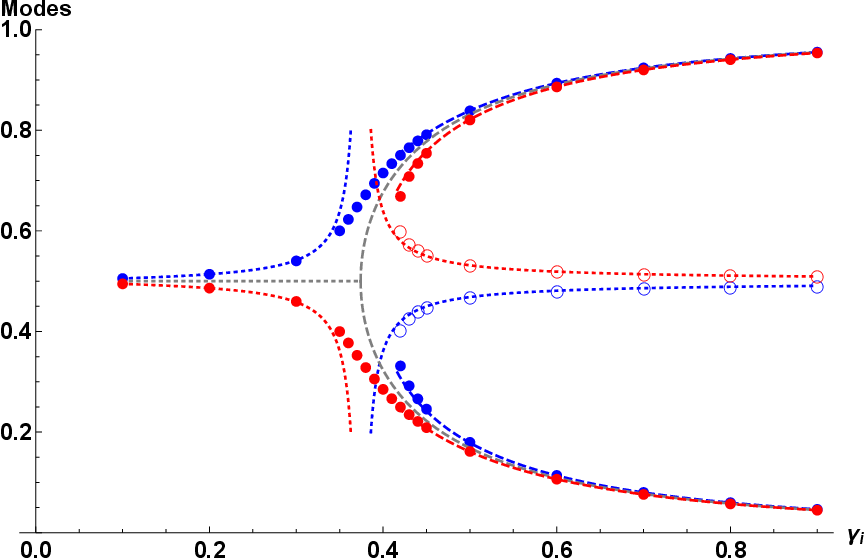}
\end{subfigure}
\begin{subfigure}{0.48\textwidth}
\caption{} 
\includegraphics[scale=0.8]{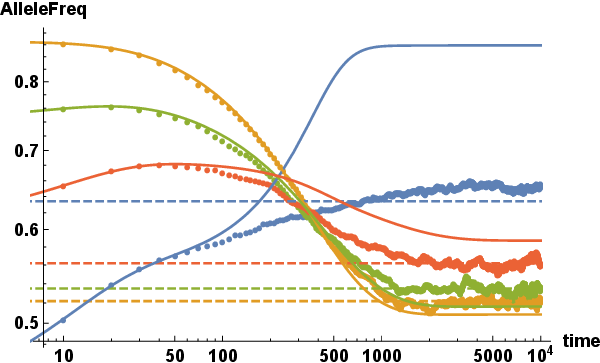}
\end{subfigure}
\caption{(a) Allele frequency at which the modes in the stationary state marginal allele frequency distribution occur  for $z_o=2$ (blue) and $-2$ (red). The points are obtained by numerically solving the cubic equation (\ref{app_mderiv}) for the modes, with closed (open) symbols denoting the maximum (minimum).   
The dotted and dashed lines, respectively, show the approximate expressions in (\ref{selmode}) and (\ref{lelmode}) for the mode frequency for finite number of loci  while 
the gray lines show (\ref{tmm}) for infinite number of loci. The parameters are the same as in Fig.~\ref{margssnz}, {\it viz.}, $N=1000, s=0.1, \mu=0.002, \bar{\gamma}=0.08$, and $L=1000$. The threshold effect obtained from the cubic equation (\ref{thfull}) and the approximation (\ref{thapproxnz}) are, respectively, $0.416$ and $0.401$, 
while   (\ref{gN}) yields ${\hat \gamma}_N(L \to \infty)=0.374$.  (b) Dynamics of deterministic allele frequency (solid lines) obtained by numerically solving (\ref{detpeqn}) and average allele frequency (points) obtained by numerically solving (\ref{Langevin}) for $5000$ independent stochastic runs for effect size $0.24$ (yellow), $0.36$ (green), $0.52$ (red), $0.8$ (blue), keeping the initial frequencies to be the same in both deterministic and stochastic model. In either model, the population is initially equilibrated to an optimum at $z_o=0$ and then the optimum is suddenly shifted to $z_o=1$ in response to which the allele frequencies evolve in time. The other parameters are $L=200, s = 0.05, \mu = 0.002, N=200$. The threshold effect ${\hat \gamma}_N(L)$ is $0.359$ and $0.493$ when the optimum is at zero and one, respectively. {The dashed lines show the average allele frequency $\av{x_i}$ in the stationary state for phenotypic optimum equal to one, and are computed using the marginal distribution (\ref{psimain}).}}
\label{peakell}
\end{figure}

It is also useful to compare these results for an infinitely large population where the deterministic allele frequency ${p_i}$ obeys \citep{Vladar:2014,Jain:2017a}
\bea
\frac{d{p_i}}{dt} =  \frac{-s {p_i} (1-{p_i}) }{2} \left[2 \gamma_i (m_1-z_o)+\gamma_i^2 (1-2 {p_i})\right]+\mu (1-2 {p_i}) \label{detpeqn}
\eea
with the deterministic trait mean, $m_1=\sum_{i=1}^L \gamma_i (2 {p_i}-1)$. Assuming that the trait mean deviation is zero in the stationary state, it has been shown that the deterministic allele frequency is in stable equilibrium below a threshold effect ${\hat \gamma}=\sqrt{ 8 \mu/s}$ and is bistable above it \citep{Vladar:2014}, and given by
\begin{subnumcases}
{{p_i^*}=\label{tmm2}}
\frac{1}{2} ~&,~$\gamma_i < {\hat \gamma}$\\
\frac{1}{2} \left(1\pm \sqrt{1-\frac{{\hat \gamma}^2}{\gamma_i^2}} \right) ~&,~$\gamma_i > {\hat \gamma}$
\label{tm2}
\end{subnumcases}
\section{Discussion}

In this article, we considered a model that describes a large but finite  population under stabilizing selection and symmetric mutation in the equilibrium state. The dynamics of adaptation following a sudden change in the phenotypic optimum of such a model have been studied  in  deterministic \citep{Vladar:2014,Jain:2015,Jain:2017a} and stochastic setting \citep{Thornton:2019,Hayward:2022,Hollinger:2023}. 

Here we  asked: in what parameter regimes the fitness epistasis can be neglected when the population is in equilibrium?  Our key result for the allele frequency distribution is summarized in \eq{epiig}. We find that for strong selection, it is sufficient to have large number of loci in order to describe the distribution without taking epistasis into account, but for weak to moderate selection, there are conditions on mutation and selection parameters given by \eq{appm_gvar} under which epistasis can be neglected. 

To connect these results at the genetic level to the phenotypic quantities, we also studied the phenotypic deviation and mean genic variance.  We find that Bulmer's argument \citep{Bulmer:1972} for mean genic variance is incorrect and therefore, their expression given by \eq{varB} is not exact, but it is a good approximation when the number of loci are large. We stress that the disagreement with Bulmer's expression is not because of linkage disequilibrium [Bulmer effect; \citep{Bulmer:1971}] as the diffusion theory  formulated here is in linkage equilibrium rather the point is that \eq{varB} neglects the epistatic effects. 
The discussion in \citet{Bulmer:1972} does not invoke the number of loci, and we are not aware of other work where this error has been pointed out. Thus, for small number of loci, as in oligogenic traits \citep{Bell:2009,Boyle:2017}, the expression (29) of \citet{Bulmer:1972} does not hold but it is a good approximation for large $L$ for any selection strength. 

Taken together, our results show that for weak to moderate selection, although genic variance matches well with Bulmer's expression, epistasis strongly affects the allele frequency distribution at loci with large effect size. Thus, while epistasis effects may not be evident at the phenotypic level, they can strongly influence the distribution at the genetic level. In a recent study by \cite{Courau:2026}, the model here has been generalized to include heterogeneous mutation rates and asymmetric mutations, and the authors arrive at the same general conclusions as in this article but via a different route.

We close the article by discussing the key differences in the stationary state of infinite \citep{Vladar:2014,Jain:2017a} and finite populations: first, it should be noted that, in general, there is no one-to-one correspondence between the multistability in a deterministic model and multimodality in the corresponding stochastic model. This point, for example, has been made in the context of biochemical reaction systems where some monostable (bistable) deterministic systems have been observed to have a bimodal (unimodal) distribution when stochastic fluctuations are allowed \citep{Hahl:2016}. Here, as \AD{${\hat \gamma}  > {\hat \gamma}_N$}, when \AD{$ {\hat \gamma}_N < \gamma_i < {\hat \gamma}$}, the deterministic allele frequencies are monostable but the allele frequency distribution is bimodal; however, outside this parameter range, the bistability (monostability) in the deterministic model corresponds to bimodality (unimodality) in the stochastic model. 

Second, the stationary state solution of the Fokker-Planck equation (\ref{LEfullfpe}) is unique [refer to Chapter 5, \citep{Kampen:1997}]; that is, it is independent of the initial allele frequencies for both small- and large-effect loci. Furthermore, as already mentioned, in a finite population, the allele frequency of a large-effect locus spends a long time (presumably,  exponentially long in population size) near one of the maxima before shifting to the other maximum. 
In contrast, in an infinite population, the stationary state allele frequency of a large-effect locus depends on the initial condition and does not shift between the two solutions given in (\ref{tm2}). These points are further illustrated in Fig.~\ref{peakell}b where the stationary state frequencies in the deterministic and stochastic model, starting from the same initial condition, are found to be quite close for small effect loci but not for the large-effect ones. For the latter case, to obtain a match between the  $N \to \infty$ limit of the stochastic model and the deterministic model, one also need to average over the initial conditions in the deterministic model.


\clearpage
\appendix

\makeatletter
\setcounter{secnumdepth}{3} 
\@addtoreset{equation}{section}
\makeatother

\setcounter{section}{0} 
\renewcommand{\thesection}{A\arabic{section}}
\renewcommand{\thesubsection}{\thesection.\arabic{subsection}}
\setcounter{equation}{0}
\renewcommand{\theequation}{\thesection.\arabic{equation}}

\section{Linkage equilibrium}
\label{app_le}


Let $f_{\sigma,\sigma^*}$ denote the frequency of a diploid individual with maternal and paternal sequences, $\sigma$ and $\sigma^*$, respectively. 
If the phenotypic effect size at locus $i$ is $\gamma_i/2$, for equivalent sexes and assuming additive phenotype-genotype map, the trait value of the individual can be written as
\bea
z &=& \sum_{i=1}^L \frac{\gamma_i}{2} (\sigma_i+\sigma^*_i)  \label{dipzdef}
\eea
with $\sigma_i=1 (-1)$ for mutant (wildtype) allele. Under random mating, the parent frequencies factorize in Hardy-Weinberg equilibrium (HWE). 
 Then the mean phenotype of the population (denoted by bar) obtained by averaging over the phenotypes of all individuals is given by 
\bea
c_1 &=& 2 \sum_{i=1}^L \frac{\gamma_i}{2} \overline{\sigma_i}=\sum_{i=1}^L \gamma_i (2 x_i-1) \label{app_c1def}
\eea
where $\overline{\sigma^*_i}=\overline{\sigma_i}$ for equivalent sexes, and $x_i$ is the frequency of the mutant allele at locus $i$. 
Similarly, one obtains the within-population (additive) genetic variance as \citep{Bulmer:1971}
\bea
C_2=\overline{z^2_{\sigma,\sigma^*}}- (\overline{z})^2 
&=& 2  \sum_{i \neq j } \frac{\gamma_i \gamma_j}{4} [\overline{\sigma_i \sigma_j}-\overline{\sigma_i} \; \overline{\sigma_j}] + 2  \sum_{i} \frac{\gamma_i^2}{4} [\overline{\sigma^2_i}-\overline{\sigma_i}^2] \label{app_genvdef}
\eea
Here, we assume that the population is in linkage equilibrium, $\overline{\sigma_i \sigma_j}=\overline{\sigma_i}\; \overline{\sigma_j}$, and obtain the (additive) genic variance, $c_2$ given by \eq{gvdef}.

\section{Diffusion theory}
\label{app_fpe}

Let $\vec{x}=\{x_1, ..., x_L\}$ denote the mutant allele frequency vector  and $P(\vec x,t)$ is the joint distribution of these frequencies. The forward Fokker-Planck equation (FPE) is then given by 
\bea
\frac{\partial P}{\partial t}= -\sum_{i=1}^L \frac{\partial}{\partial x_i} [\av{\delta x_i} P]+\frac{1}{2} \sum_{i,j}\frac{\partial^2}{\partial x_{i} \partial x_{j}} [\av{\delta x_{i} \delta x_{j}} P] \label{LEfullfpe}
\eea
where, the mean, variance and covariance in the change in the allele frequencies due to stabilizing selection, mutation and random genetic drift are described below. 

First consider the effect of selection: Since $x_i=\sum_{\sigma, \sigma} \delta_{\sigma_i,1} f_{\sigma,\sigma^*}$, the (deterministic) change in the mutant allele frequency after selection is given by \citep{Kimura:1964,Bulmer:1972}
\bea
\av{\delta x_i}_s =-\frac{s}{2} [2 \gamma_i x_i (1-x_i)(\bar z-z_o)+\gamma_i^2 x_i (1-x_i) (1-2 x_i)]
\eea
on using that the population is in HWE before selection. For symmetric mutation rates, the deterministic change in $x_i$ due to mutation is given by 
\be
\av{\delta x_i}_m=\mu (1-x_i)-\mu x_i=\mu (1-2 x_i)
\ee
Thus the change $\av{\delta x_i}$ due to selection and mutation is obtained by adding their respective contributions, and given by 
\bea
\av{\delta x_i} &=&  -\frac{s}{2} [2 \gamma_i x_i (1-x_i) \Delta c_1+  \gamma_i^2 x_i (1-x_i) (1-2 x_i)] 
+ \mu (1-2 x_i) \label{LEdetp}
\eea
Note that $\av{\delta x_i}$ depends on the frequency of other loci also through $\Delta c_1=\sum_{j=1}^L \gamma_j (2x_j-1)-z_o$.  
Finally, the change due to multinomial sampling in the Wright-Fisher process is given by 
\be
\av{\delta x_{i} \delta x_{j}} =\frac{x_i(1-x_i)}{2 N} \delta_{ij} \label{varx}
\ee
where the covariance vanishes as the population is in linkage equilibrium. 

In the steady state [refer to Chapter 6, \citep{Risken:1996}; \citep{Kimura:1964}], the LHS of  \eq{LEfullfpe} is zero, and therefore, the divergence of the total probability current vanishes: $\sum_{i=1}^L \frac{\partial J_i(\vec x)}{\partial x_i}=0$, where
\bea
J_i &=&\av{\delta x_i} P-\frac{1}{4 N} \frac{\partial }{\partial x_{i}} [x_{i} (1-x_i)P] 
\eea
In general, it is not necessary that each $J_i$ be same for all loci, and moreover, equal to zero.  But if we do assume that $J_i=0$ for all $i$, we obtain 
\bea
\frac{\partial \ln P}{\partial x_{i}}=\frac{4 N \av{\delta x_i}-(1-2 x_i)}{x_i (1-x_i)} \equiv F_i(\vec x)
\eea
Then, for $\Phi=\ln P$, the total derivative 
$d \Phi=\sum_i \frac{\partial \Phi}{\partial x_{i}} dx_i$ exists iff $\frac{\partial F_i}{\partial x_{j}} = \frac{\partial F_j}{\partial x_{i}}$; using \eq{LEdetp}, we verify that this condition is indeed satisfied, and therefore, 
%
we can write
\bea
\ln P(\vec x) &=&  \sum_{i=1}^L \int dx_i \frac{4 N \av{\delta x_i}-(1-2 x_i)}{x_i (1-x_i)}+e^C
\eea
 We thus obtain \cite{}
\bea
P(\vec x)
&=& C \; e^{-{N s (c_1-z_o)^2}}  e^{-2 N s \sum \gamma_i^2 x_i (1-x_i)}  \prod_{i=1}^L [x_i (1-x_i)]^{4 N\mu-1} \label{LEPss}
\eea
where, the constant $C$ is determined using the normalization condition, $\int_0^1 \prod_{i=1}^L dx_i \; P(\vec x)=1$.


\section{Marginal distribution}
\label{app_margdistr}

The marginal distribution $\psi(x_i)$ of the allele frequency in the stationary state can be found by integrating the joint distribution $P(\vec x)$ given by (\ref{LEPss1}) over the frequencies of all but the $i$th locus; this gives 
\bea
\psi(x_i) &\propto& \psi_B(x_i) \int_0^1... \int_0^1 e^{-Ns (c_1-z_o)^2} \prod_{j\neq i}  dx_j \psi_B(x_j)
\label{app_jmarg}
\eea
where 
\be
\psi_B(x_i) \propto e^{-2Ns\gamma_i^2 x_i(1-x_i)}(x_i(1-x_i))^{4N\mu-1}
\ee
and $c_1=\sum_{j=1}^L \gamma_j (2 x_j-1)$. To evaluate the multiple integrals in (\ref{app_jmarg}), we take $X'=\sum_{j \neq i} \gamma_j (2 x_j-1)$ and rewrite its RHS as 
\bea
\frac{\psi(x_i)}{\psi_B(x_i)}\propto \int_{-\Gamma'}^{\Gamma'} dX' e^{-N s (X'+\gamma_i (2 x_i-1)-z_o)^2} 
\times \int_0^1...\int_0^1  \delta[X'-\sum_{j \neq i} \gamma_j (2 x_j-1)] \prod_{j \neq i}  dx_j \psi_B(x_j)\label{cltapp}
\eea
where $\Gamma'=\sum_{j \neq i} \gamma_j$, and $X'$ is a sum of independent but non-identically distributed random variables chosen from $\psi_B(x_j)$. 
The inner integrals on the RHS of the above equation can be calculated by appealing to the central limit theorem for large $L$, and we get
\bea
\frac{\psi(x_i)}{\psi_B(x_i)} &\stackrel{L \gg 1}{\propto}& \int_{-\infty}^\infty dX' e^{-N s (X'+\gamma_i (2 x_i-1)-z_o)^2} e^{-\frac{(X'-\kappa_{1,i})^2}{2 \kappa_{2,i}}} \\
&{\propto}& \exp \left[ -\frac{N s  (\gamma_i (2 x_i-1)-z_o)^2}{1+2 N s \kappa_{2,i}}\right]\label{margclt}
\eea
where $\kappa_{1,i}$ and $\kappa_{2,i}$ are, respectively, the mean and variance of the sum $X'=\sum_{j \neq i} \gamma_j (2 x_j-1)$ when averaged over the (normalized) distribution $\prod_{j \neq i} \psi_B(x_j)$.

Since $\psi_B(x_i)$ is symmetric about $x_i=1/2$, it follows that the mean $\kappa_{1,i}=0$.
In the marginal distribution $\psi(x_i)$, the effect of other $L-1$ loci due to epistatic interactions in the phenotypic fitness function appears through the variance $\kappa_{2,i}$. {Since {$\av{x_i}_{\psi_B}=1/2$} for all loci, the variance $\kappa_{2,i}=\sum_{j \neq i} 4 \gamma_j^2 (\av{x_j^2}_{\psi_B}-\av{x_j}_{\psi_B}^2)$ which shows that $\kappa_{2,i}$ is simply a weighted sum of the variance of the distribution $\psi_B(x_i)$, and is given by
\bea
\kappa_{2,i} &=& \sum_{j=1, j \neq i}^L  \frac{\gamma_j^2}{1+8 N \mu}  \frac{{_1}F_1(\frac{3}{2},4 N \mu+\frac{3}{2}, \frac{N s \gamma_j^2}{2})}{{_1}F_1(\frac{1}{2},4 N \mu+\frac{1}{2}, \frac{N s \gamma_j^2}{2})}  \label{app_k2i}
\eea
where ${_1}F_1(a, b, z)$ is the confluent hypergeometric function. 

\section{Statistical variance in the absence of epistasis}
\label{app_sve}


{Here, we study the properties of $\kappa_2=\sum_{j=1}^L 4 \gamma_j^2 (\av{x_j^2}_{\psi_B}-\av{x_j}_{\psi_B}^2)$. Note that $\kappa_2$ is the standard statistical variance while the mean genic variance $\av{c_2}_{\psi_B}$ is related to the expectation of $x (1-x)$, although both are calculated w.r.t. $\psi_B$. We have 
\bea
\kappa_2 &=& \sum_{j=1}^L 4 \gamma_j^2  \left(\frac{1}{4 (1+8 N \mu)}  \frac{{_1}F_1(\frac{3}{2},4 N \mu+\frac{3}{2}, \frac{N s \gamma_j^2}{2})}{{_1}F_1(\frac{1}{2},4 N \mu+\frac{1}{2}, \frac{N s \gamma_j^2}{2})} \right)\\
&\approx &  L \int_0^\infty  \frac{p(\gamma) \gamma^2}{1+8 N \mu}  \frac{{_1}F_1(\frac{3}{2},4 N \mu+\frac{3}{2}, \frac{N s \gamma^2}{2})}{{_1}F_1(\frac{1}{2},4 N \mu+\frac{1}{2}, \frac{N s \gamma^2}{2})} \label{app_svar}
\eea
where $p(\gamma)$ is the distribution of effects. 
For exponentially-distributed effect size with mean ${\bar \gamma}$, we have
\bea
\kappa_2 &\approx & \frac{L {\bar \gamma}^2}{1+8 N \mu} \int_0^\infty  \frac{{_1}F_1(\frac{3}{2},4 N \mu+\frac{3}{2}, \frac{N s {\bar \gamma}^2}{2} y^2)}{{_1}F_1( \frac{1}{2},4 N \mu+\frac{1}{2}, \frac{N s {\bar \gamma}^2}{2} y^2)} y^2  e^{-y} dy \label{app_svarint}
\eea
As the above integral does not appear to be exactly solvable, we estimate it by noting that the ratio of the confluent hypergeometric function in the above integrand is a monotonically increasing function of its argument, and for {{\it fixed} $U=4 N \mu$}, this ratio may be approximated using (13.7.1) and (13.11.1) of \citet{DLMF} as 
\bsn
{\frac{{_1}F_1(\frac{3}{2}, U+\frac{3}{2}, a)}{{_1}F_1( \frac{1}{2},U+\frac{1}{2}, a)} \approx \label{Fratap}}
1+\frac{4 U a}{4 U^2+8 U+3} ~,&~ $a \ll 1$ \\
(2 U+1) \left(1 -\frac{U}{a} \right) ~,&~ $a \gg 1$
\esn
Thus, for a given $4N\mu$, using (\ref{Fratap}) in (\ref{app_svarint}), we obtain
 \bsn
{\kappa_2 \approx \label{app_gvar}}
\frac{L{\bar \gamma}^2}{1+8 N \mu} \left(2+\frac{1536 {\bar \gamma}^2  N^2 \mu ^2}{{\hat \gamma}^2 (8 N \mu
   +1) (8 N \mu +3)}\right)  &,~$N s {\bar \gamma}^2 \ll 2 $ \label{app_gvar1} \\
L {\bar \gamma}^2  \left(2-\frac{{\hat \gamma}^2}{{\bar \gamma}^2}\right)&,~ $N s  {\bar \gamma}^2 \gg 2$ \label{app_gvar2}
\esn
where, ${\hat \gamma}=\sqrt{\frac{8 \mu}{s}}$. 

When selection is weak ($N s  {\bar \gamma}^2 \ll 2$) {and mutation is strong ($4 N \mu \gg 1$)}, 
(\ref{app_gvar1}) shows that $\kappa_2$ approaches zero as $1/N$; in contrast, when selection is {stronger than mutation, 
due to (\ref{app_gvar2}), the variance $\kappa_2 \to 2 L {\bar \gamma}^2$} so that it remains nonzero in the deterministic limit. 
 These results can be understood as follows: as the width of $\psi_B(x_i)$ about a maximum is expected to decrease  with $N$, 
 for large $N$, one may approximate $\psi_B(x_i)$ by a Dirac delta function centred at $1/2$ for small effect locus, and an average of two Dirac delta functions located at $p_{i,\pm}=\frac{1}{2} \left(1 \pm \sqrt{1-\frac{{\hat \gamma}^2}{\gamma_i^2}} \right)$ for large-effect locus \citep{Vladar:2014}.  Thus, for small-effect locus, as the distribution is unimodal and sharply-peaked in large populations, the variance vanishes in the deterministic limit. But for large-effect locus, as a consequence of the bimodality, the distribution remains broad resulting in a nonzero variance.  

\section{Corrections to central limit theorem}
\label{app_edge}

Consider the distribution of the random variable, $Y=\frac{\sum_{j=1}^L \gamma_j (2 x_j-1)}{\sqrt{\kappa_2}}$ where  $x_i$'s are independently distributed according to the (normalized) distribution, $\psi_B(x_i) \propto e^{-2Ns\gamma_i^2 x_i(1-x_i)}(x_i(1-x_i))^{4N\mu-1}$ and $\kappa_2=\sum_{j=1}^L \av{ \gamma_j^2 (2 x_j-1)^2}_{\psi_B}$ where we have used that the mean $\av{\gamma_j (2 x_j-1)}_{\psi_B}=0$. Then, from (43) of  \citet{Blinnikov:1998}, we have
\be
\textrm{Prob}(Y) \propto e^{-\frac{Y^2}{2}} \left[ 1+ \frac{\kappa_4}{\kappa_2^2} \frac{3-6 Y^2+Y^4}{24}+{\cal O}(L^{-2})\right] \label{edgedist}
\ee
where, $\kappa_4=\sum_{j=1}^L \av{\gamma_j^4 (2 x_j-1)^4}_{\psi_B} -3 \av{ \gamma_j^2 (2 x_j-1)^2}^2_{\psi_B}$ is the fourth cumulant of the random variable $\gamma_j (2 x_j-1)$  w.r.t. the (normalized) distribution $\psi_B(x_j)$. Using the above result in (\ref{cltapp}) for $X'=\sqrt{\kappa_2} Y-\gamma_i (2 x_i-1)$ and performing the integral over $X'$, we arrive at 
\bea
\frac{\psi(x_i)}{\psi_B(x_i)} \propto 1+\frac{\kappa_{4}-4 \kappa_2 (\gamma_i (2 x_i-1)-z_o)^2}{8 \kappa_2^2} +{\cal O}(L^{-2})\label{margsz0}
\eea

\section{Stationary state genic variance}
\label{app_var}


Consider the  genic variance \eq{gvdef} when it is averaged over the stationary state distribution (\ref{margsz0}). For $z_o \ll L$, we can write
\bea
\av{x_i (1-x_i)} &=& \int_0^1 dx_i x_i (1-x_i) \psi(x_i) \\
&\approx& \frac{\int_0^1  dx_i x_i (1-x_i) \psi_B(x_i) \left( 1-\frac{(\gamma_i (2 x_i-1)-z_o)^2}{2 (\kappa_2+\frac{\kappa_4}{8 \kappa_2})}+...\right)}{\int_0^1 dx_i \psi_B(x_i) \left( 1-\frac{(\gamma_i (2 x_i-1)-z_o)^2}{2  (\kappa_2+\frac{\kappa_4}{8 \kappa_2})}+...\right)}  \\
&=&  \frac{\int_0^1  dx_i x_i (1-x_i) \psi_B(x_i)}{\int_0^1 dx_i \psi_B(x_i)} \nonumber \\
&+&  \frac{1}{2 (\kappa_2+\frac{\kappa_4}{8 \kappa_2})} \frac{\int_0^1  dx_i x_i (1-x_i) \psi_B(x_i) \int_0^1 dx_i \psi_B(x_i) (\gamma_i^2 (2 x_i-1)^2+z_o^2)}{[\int_0^1 dx_i \psi_B(x_i)]^2} \nonumber \\
&-& \frac{1}{2 (\kappa_2+\frac{\kappa_4}{8 \kappa_2})} \frac{\int_0^1  dx_i x_i (1-x_i) \psi_B(x_i) (\gamma_i^2 (2 x_i-1)^2+z_o^2)}{\int_0^1 dx_i \psi_B(x_i)} 
\eea
On summing over all the loci, the last equation gives the mean genic variance to be 
\bea
&&\frac{\av{c_2}}{L}= \frac{U}{1+2 U}  \int_0^\infty d\gamma  \gamma^2 p(\gamma) \frac{{\cal F}(\frac{3}{2},\gamma)}{{\cal F}(\frac{1}{2},\gamma)} \nonumber \\ 
&+&\frac{1}{(\kappa_2+\frac{\kappa_4}{8 \kappa_2})} \frac{U}{1+2 U} \int_0^\infty d\gamma  \gamma^4 p(\gamma)   \left( \frac{U+1}{2 U+3}\frac{{\cal F}(\frac{5}{2},\gamma)}{{\cal F}(\frac{1}{2},\gamma)} - \frac{U}{2 U+1}\frac{{\cal F}^2(\frac{3}{2},\gamma)}{{\cal F}^2(\frac{1}{2},\gamma)}  \right) 
\label{statvar}
\eea
where, for brevity, we have defined ${\cal F}(\alpha,\gamma)={_1}F_1\left(\frac{1}{2},4 N \mu+\alpha, \frac{N s \gamma^2}{2} \right)$ and $U=4 N \mu$. 

The first term on the RHS of (\ref{statvar}) is independent of $z_o$ and is equal to $\av{c_2}_{\psi_B}$, and the second term that gives the $L$-dependent correction is of order $L^{-1}$ and is also seen to be {independent} of the phenotypic optimum. Thus, when $z_o \ll L$, the mean genic variance (to ${\cal O}(L^{-1}))$ does not depend on the location of the phenotypic optimum.

\section{Threshold effect size and mode frequency} 
\label{app_modes}

On setting the derivative $\frac{d \psi}{dx_i}$ of the marginal distribution \eq{psimain} equal to zero, we arrive at  the following cubic equation for the modes $\chi_i(L)$:
\be
(4 N \mu-1) \frac{(1-2 \chi_i)}{ \chi_i (1-\chi_i)}-2 N s \gamma_i^2 (1-2 \chi_i)+\frac{4  N s \gamma_i(z_o-{\gamma_i (2 \chi_i-1)})
}{1+2 N s \kappa_{2,i}}=0 \label{app_mderiv}
\ee
The above equation has two complex roots and one real root below the threshold effect ${\hat \gamma}_N(L)$ and three real roots above it. This change in the behavior of $\chi_i(L)$ occurs when the discriminant of the above cubic polynomial {[(1.11.12), \citep{DLMF}]} is equal to zero: 
\be
D=a_2^2 a_1^2-4 a_3 a_1^3+18 a_0 a_2 a_3 a_1-a_0 \left(4 a_2^3+27 a_0 a_3^2\right)=0
\label{thfull}
\ee
where $a_i, i=0,1,2,3$ are the coefficients of $x^i$ in (\ref{app_mderiv}). The resulting equation is a 3rd order equation in ${\hat \gamma}^2_N(L)$ and can be solved numerically to obtain the threshold effect for finite $N$ and $L$. 

One can, however, obtain an analytical expression for ${\hat \gamma}_N(L)$ when the number of loci are very large. Since $\kappa_{2}$ grows linearly with $L$, for infinite number of loci, the last term on the LHS of (\ref{app_mderiv}) vanishes yielding (\ref{gN}) and (\ref{tmm}) for the threshold effect ${\hat \gamma}_N$ and mode allele frequency $\chi_i$, respectively, which are independent of $z_o$. 
For large $L$, using (\ref{margclt}), we find that the steady state distribution has maximum at
\bsn
{\chi_i(L)=\label{tmmell}}
\frac{1}{2}+\frac{1}{2 N s \kappa_2} \frac{\gamma_i z_o}{{\hat \gamma}_N^2-\gamma_i^2}~&,~$\gamma_i < {\hat \gamma}_N(L)$ \label{selmode}\\
\frac{1}{2} \left(1 \mp \sqrt{1-\frac{{\hat \gamma}_N^2}{\gamma_i^2}} \right)+\frac{1}{2 N s \kappa_2} \frac{{\hat \gamma}_N^2(z_o \pm \sqrt{\gamma_i^2-{\hat \gamma}_N^2})}{2 \gamma_i (\gamma_i^2-{\hat \gamma}_N^2)} ~&,~$\gamma_i > {\hat \gamma}_N(L)$ \label{lelmode}
\esn
where ${\hat \gamma}_N(L)$ denotes the threshold effect for finite $L$. For $\gamma_i > {\hat \gamma}_N(L)$, the minimum in the allele frequency is given by the expression on the RHS of  (\ref{selmode}).

As shown in Fig.~\ref{peakell}, a threshold effect exists below which (\ref{app_mderiv}) has only one real root and corresponds to the maximum in the unimodal distribution. But, above the threshold effect, two additional real roots of (\ref{app_mderiv}) appear which give the allele frequency at which the minimum and the second maximum of the bimodal distribution occur. Thus, at the threshold effect, for positive (negative) $z_o$, the minimum and the low-frequency (high-frequency) maximum of the bimodal distribution coincide. On matching (\ref{selmode}) and {the relevant solution in} (\ref{lelmode}), we get
\be
\pm 2 N s \kappa_2 \Delta^3=3 {\hat \gamma}_N^2 z_o \pm {\hat \gamma}_N^2 \Delta  +2 z_o \Delta^2
\label{deapprox}
\ee
where $\Delta^2={\hat \gamma}^2_N(L)-{\hat \gamma}^2_N$ decreases with increasing $L$. The above cubic equation for $\Delta$ is exactly solvable, and has two complex roots and one real root. Here, we estimate the real root by noting  that the first term on the RHS which is independent of $L$ can be balanced if $\kappa_2 \Delta^3$ is also independent of $L$ thus yielding 
\be
{\hat \gamma}_N(L) \approx {\hat \gamma}_N+\frac{1}{2 {\hat \gamma}_N} \left(\frac{3 {\hat \gamma}_N^2 |z_o|}{2 N s \kappa_2} \right)^{2/3} +{\cal O}(L^{-1})~,~z_o \neq 0
\label{thapproxnz}
\ee
which shows that the deviation ${\hat \gamma}_N(L)-{\hat \gamma}_N$ decays rather slowly with $L$. 
The above expression also shows that the threshold effect always increases with the absolute value of the phenotypic optimum. But it does not change if the phenotypic optimum shifts between $z_o$ and $-z_o$. The threshold effect is also larger when selection is weaker ($N s {\bar \gamma}^2 \ll 2$) or the quantitative trait is controlled mostly by small-effect loci. For phenotypic optimum at zero, (\ref{deapprox}) gives 
\be
{\hat \gamma}_N(L) \approx {\hat \gamma}_N+\frac{{\hat \gamma}_N}{4 N s \kappa_2}
\label{thapproxz}
\ee
so that the deviation  ${\hat \gamma}_N(L)-{\hat \gamma}_N$ is of order $L^{-1}$. 

For small $z_o$ and large $L$, (\ref{thapproxnz}) shows that the threshold effect does not differ much from the infinite-loci result (\ref{gN}) when selection is strong ({$N s {\bar \gamma}^2 \gg 2$}). 
Furthermore, (\ref{thapproxnz}) predicts that the threshold effect for large but finite number of loci is always larger than ${\hat \gamma}_N$, 
and increases with the magnitude of the phenotypic optimum. Thus, a locus classified as a large-effect locus for a phenotypic optimum at zero can become a small-effect locus if the phenotypic optimum is large enough.
 This is because for large, positive (negative) phenotypic optimum, population will be well-adapted if the $+$ ($-$) allele's frequency at most loci is close to fixation (that is, the distribution is unimodal and heavily skewed towards high frequency).

\clearpage

\printcredits

\bibliographystyle{cas-model2-names}
\bibliography{/Users/jain/Desktop/PG_papers/master_evolAM.bib,/Users/jain/Desktop/PG_papers/master_evolNZ.bib}

\clearpage

%
%
%

\end{document}